\documentclass{aa}  
\usepackage{graphicx}
\usepackage{subfigure}
\usepackage{xcolor}
\usepackage{txfonts}

\usepackage[colorlinks=true,linkcolor=blue,citecolor=blue]{hyperref}
\begin{document}

%   \title{Depicting the convection cells in the (sur)face of the Mira star R~Car\thanks{Based on observations made with the Very Large Telescope Interferometer (VLTI) at the Paranal Observatory under programs ID 0104.D-0390(A), 0104.D-0390(B), 0104.D-0390(C), and 60.A-9237(A).}}

\title{A new dimension in the variability of AGB stars: 
convection patterns size changes with pulsation\thanks{Based on observations made with the Very Large Telescope Interferometer (VLTI) at the Paranal Observatory under programs ID 0104.D-0390(A), 0104.D-0390(B), 0104.D-0390(C), and 60.A-9237(A).}}

   \author{A. Rosales-Guzmán \inst{1,2},
        J. Sanchez-Bermudez \inst{1,3},
        C. Paladini \inst{2},
        B. Freytag\inst{4},
        M. Wittkowski \inst{5},
         A. Alberdi\inst{6},
         F. Baron\inst{7},
         J.-P. Berger\inst{8},
         A. Chiavassa\inst{9},
         S. Höfner\inst{4},
         A. Jorissen\inst{10},
         P. Kervella\inst{11},
         J.-B. Le Bouquin\inst{8},
         P. Marigo\inst{12},
         M. Montargès\inst{11},
         M. Trabucchi\inst{12},
         S. Tsvetkova\inst{13},          
         R. Schödel\inst{6},
         S. Van Eck\inst{10}
          }

   \institute{Instituto de Astronomía, Universidad Nacional Autónoma de México, Apdo. Postal 70264, Ciudad de México, 04510, México\\ %1
              \email{jarosales@astro.unam.mx}
        \and
             European Southern Observatory (ESO), Alonso de Córdova 3107, Vitacura, Santiago, Chile %2
        \and
            Max-Planck-Institut f\"ur Astronomie, K\"{o}nigstuhl 17, D-69117 Heidelberg, Germany
        \and
            Theoretical Astrophysics, Department of Physics and Astronomy, Uppsala University, Box~516, SE-751~20 Uppsala, Sweden %3
        \and
            European Southern Observatory (ESO), Karl-Schwarzschild-Str. 2, D-85748 Garching bei München, Germany %4
       \and
            Instituto de Astrofísica de Andalucía, Glorieta de la Astronomía s/n, 18008 Granada, España %5
       \and
            Center for High Angular Resolution Astronomy and Department of Physics and Astronomy, Georgia State University, P.O. Box 5060, Atlanta, GA 30302-5060, USA %6
       \and 
       UJF-Grenoble 1/CNRS-INSU, Institut de Planétologie et d’Astrophysique de Grenoble (IPAG) UMR 5274, Grenoble, France %7
       \and 
            Université Côte d’Azur, Observatoire de la Côte d’Azur, CNRS, Laboratoire Lagrange, France %8
       \and
            Institut d’Astronomie et d’Astrophysique, Université Libre de Bruxelles, CP 226, Boulevard du Triomphe, 1050 Brussels %9
       \and
            LESIA, Observatoire de Paris, Université PSL, CNRS, Sorbonne Université, Université Paris Cité, 5 place Jules Janssen, 92195 Meudon, France %10
       \and
            Dipartimento di Fisica e Astronomia Galileo Galilei, Universita` di Padova, Vicolo dell’Osservatorio 3, I-35122 Padova, Italy %11
       \and
            Institute of Astronomy and NAO, Bulgarian Academy of Sciences, 72 Tsarigradsko shose, 1784 Sofia, Bulgaria %12
             }

   \date{}

% \abstract{}{}{}{}{} 
% 5 {} token are mandatory
 
  \abstract
   {Stellar convection plays an important role in atmospheric dynamics, wind formation and the mass-loss processes in Asymptotic Giant Branch (AGB) stars. However, a direct characterization of convective surface structures in terms of size, contrast, and life-span is quite challenging. Spatially resolving these features requires the highest angular resolution.}
   {We aim at characterizing the size of convective structures on the surface of the O-rich AGB star \object{R Car} to test different theoretical predictions, based on mixing-length theory from solar models.}
   {We used infrared low-spectral resolution (R$\sim$35) interferometric data in the $H-$band ( $\sim$ 1.76 $\mu$m) with the instrument PIONIER at the Very Large Telescope Interferometer (VLTI) to image the star's surface at two epochs separated by $\sim$ 6 years. Using a power spectrum analysis, we estimate the horizontal size of the structures on the surface of R Car. The sizes of the stellar disk, at different phases of a pulsation cycle, were obtained using parametric model-fitting in the Fourier domain. }
   {Our analysis supports that the sizes of the structures in R Car are correlated with variations of the pressure scale height in the atmosphere of the target, as predicted by theoretical models based on solar convective processes. We observe that these structures grow in size when the star expands within a pulsation cycle. While the information is still scarce, this observational finding highlights the role of convection in the dynamics of those objects. New interferometric imaging campaigns with the renewed capabilities of the VLTI are envisioned to expand our analysis to a larger sample of objects. }
{}


    \titlerunning{R Car convective structures seen with PIONIER-VLTI}
    \authorrunning{Rosales-Guzm\'an, et al.}
   \maketitle
%
%-------------------------------------------------------------------

\section{Introduction}
%The asymptotic giant branch (AGB) phase is one of the final stages in the evolution of low-to-intermediate mass stars. 

After a long life ($\sim$  10$^9$ - 10$^{10}$ years) on the main-sequence, low- to intermediate-mass stars (M $\le$ 8 M$_{\odot}$) evolve into the Asymptotic Giant Branch (AGB) phase. During this evolutionary stage, their diameters grow to be up to several hundred times larger than their main-sequence ones, while their surface temperatures drop to T $\sim$ 2000 - 3000 K. Hence, they become redder in the Hertzsprung-Russell (H-R) diagram.
%\citep[see ][and references therein]{hofner2018mass}. 

\begin{figure*}[h!]
    \centering
    \subfigure[]{\includegraphics[width=8cm]{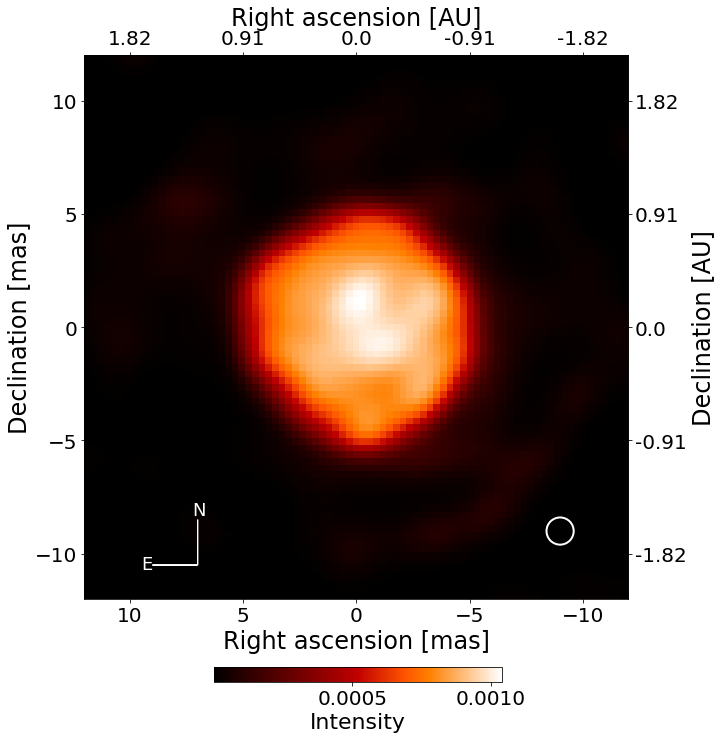}}
    \subfigure[]{\includegraphics[width=8cm]{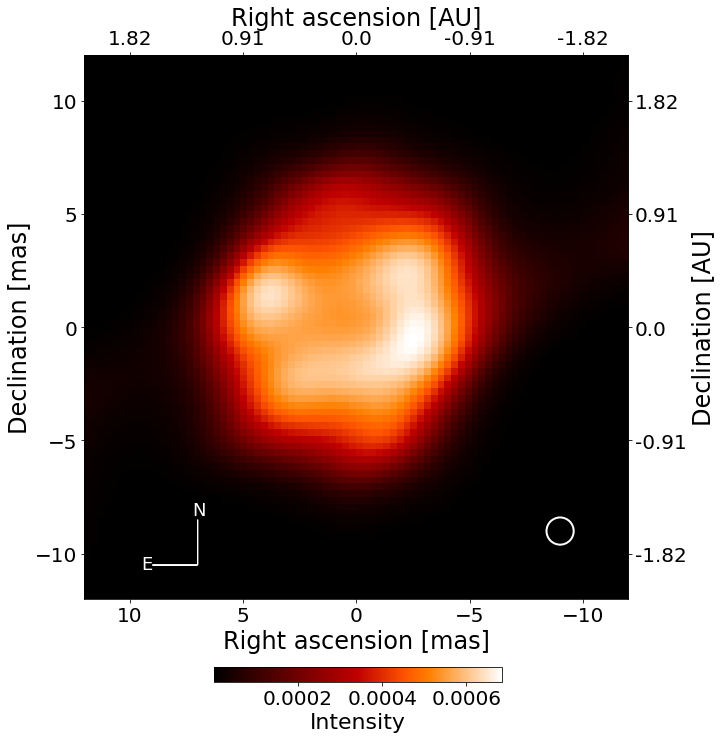}}
    \caption{Best-fit reconstructed images for the 2014 (left) and 2020 (right) datasets. The white circles located in the right corner of the images correspond to the maximum resolution  of the interferometer.}
    \label{fig:best_mean_imgs}
\end{figure*}

The late phases of stellar evolution are largely affected by mass loss due to convection and pulsation processes, which in turn, significantly affect the structure and evolution of stars \citep[][]{kupka2004convection}. 
In the case of convection, despite its importance, our current knowledge is still limited. Observationally, we mainly rely on studying the convective cells on the surface of the Sun. \citet{Schwarzschild_1975} suggested that the convection length scale is proportional to the pressure scale height of a stellar atmosphere. Hence, while the Sun has a couple of million cells on its surface, AGB stars will have (up to) a few tens of them. However, testing this hypothesis directly on resolved stellar surfaces has not been possible until very recently. More recent mixing-length theoretical models \citep[e. g., ][]{freytag1997scale, trampedach2013grid, tremblay2013granulation} establish a relation between the convection length and stellar parameters such as effective temperature or gravity for different stars across the H-R diagram.

High-angular resolution techniques are necessary to resolve those structures and infrared (IR) interferometers played an important role in the last decade measuring the diameters of stellar disks and later on resolving smaller scale structures, which were related to convection \citep[see e.g.,][]{LeBouquin_2011, wittkowski2017aperture, Perrin_2020}. 
One of the main challenges of studying convection in AGB stars is that their atmospheres are composed of molecular gases (e.g., H$_2$O, CO, HCN, etc.) and dust, obscuring the photosphere and making it difficult to observe the convective patterns. This is particularly important for carbon-rich AGBs as shown in the $H$-band images by \citet{wittkowski2017aperture}. Oxygen-rich AGBs are therefore better-suited targets to study convection because the dust is mostly transparent at the $H$ ($\lambda_0$ = 1.76 $\mu$m) and $K$ ($\lambda_0$ = 2.2 $\mu$m) bands, thus, allowing us to probe the photosphere observationally.

\citet{paladini2018large} used the PIONIER \citep[Precision Integrated-Optics Near-infrared Imaging ExpeRiment;][]{LeBouquin_2011} instrument to image in the $H-$band the stellar disk of the AGB star $\pi^1$ Gru. The interferometric images revealed the presence of bright spots that were associated with convective patterns. Those authors characterized their size to compare it, for the first time, with predictions from mixing-length theoretical models \citep[e. g., ][]{freytag1997scale, trampedach2013grid, tremblay2013granulation}. This reveals that the observational size of the convective structures agrees with a scale proportional to $\sim$10 times the pressure scale height of the atmosphere. \citet{climent2020vlti} and \citet{norris2021long} conducted similar comparisons for a couple of red supergiant (RSG) stars. However, the estimated size of the convective patterns in RSG stars did not follow the theoretical predictions extrapolated from the Sun's convection.

The target of this study is \object{R Car}, a M-type Mira star that is bright in the near-infrared and is located at 182 $\pm$ 16 pc from Earth \citep{brown2021gaia}. The light curve of this source exhibits an amplitude of 7.4 mag in the $V$-band and of -1.23 mag in the $K$-band \citep{whitelock2008agb} with a period of 314 days \citep{samus2017general, lebzelter2005study, mcdonald2012fundamental}. Its light curve is reasonably sinusoidal and shows no obvious signature of a (superimposed) secondary period. \citet{mcdonald2012fundamental} derived an effective temperature, $T_\mathrm{eff}$ = 2800~K, bolometric luminosity, $L$ = 4164 L$_\odot$, and period $P$ = 314 days for \object{R Car} by comparing BT-Settl model atmospheres \citep{allard2003model} to Spectral Energy Distributions (SEDs) created from different surveys spanning wavelengths from 420 nm to 25 $\mu$m, including data from Hipparcos, Tycho, The Sloan Digital Sky Survey (SDSS), among others. \citet{groenewegen1999millimeter} found a mass-loss rate  $\mathrm{M_\odot} <$ 1.6 $\times$ 10$^{-9}$ for R Car trough model fitting of the SED in the near infrared (NIR) and from the observed CO (3-2) line intensities. \citet{takeuti2013method} employed the relation between the period and the mean density of a pulsating star \citep[see e.g.][]{cox2017theory} to derive a mass of $\mathrm{M}$ = 0.87 M$_\odot$ for this source.
%\object{R Car} has a $T_\mathrm{eff} = 2800$~K, a luminosity of L = 4164 L$_\odot$ \citep{mcdonald2012fundamental}, a mass of $M$ = 0.87 M$_\odot$ \citep{takeuti2013method}, and the mass-loss rate is less than 1.6 $\times$ 10$^{-9}$ M$_\odot$~yr$^{-1}$ \citep{groenewegen1999millimeter}.  

The photosphere of R Car was imaged with PIONIER, as part of the \textit{"Interferometric Imaging Contest"} \citep{monnier20142014}.  The data used by those authors were obtained in January 2014 at the pulsation phase $\phi = 0.78 \pm 0.02$, where the pulsation phases of R Car are defined as $\phi = (t - t_0)/P$, with $t$ being the mean observing epoch, calculated as the median epoch between the start and end of the observations, $t_0$ being the reference epoch corresponding to the first maximum of the light curve, and $P$ being the period of the star. These phases were obtained from the visual light-curve of the object (see Fig. \ref{fig:light_curves}). The reconstructed image shows a disk with an angular diameter of about 10 mas ($\sim$1.9 au), and a couple of bright spots interpreted as convective cells. No further analysis has been conducted out on those images until now.

%The data were obtained in January 2014 at the pulsation phase\footnote{throughout the text we will refer to the pulsation phases of \object{R Car} obtained from the Visual light-curve of the object (see Fig.\ref{fig:light_curves}), and they are defined as $\phi$ = (t-t$_0$)/T where t is the observing epoch, t$_0$ is a reference epoch, and T the period of the star.} $\phi = 0.78 \pm 0.02$.

More recently, \citet{Rosales-Guzman_2023} used GRAVITY-VLTI  \citep{2017A&A...602A..94G} data to image this source in the $K-$band. This study reveals changes in the angular size of the target from 14.84 $\pm$ 0.06 mas to 16.67 $\pm$ 0.05 mas ($\sim$ 2.7 - 3 au) at two different phases ($\phi$ = 0.56 and 0.66) within the same pulsation period. The detected CO inner-most layers appear to have a clumpy structure and their elevation  ($\sim$ 2.3 R$_*$) above the stellar disk is consistent with a dust formation region that favors Magnesium composites  \citep[Mg$_2$SiO$_4$ and MgSiO$_3$; see e.g.][]{Bladh_2012, hofner_2022} as the origin of dust-driven winds in M-type AGB stars.

In this work, we analyze the 2014 PIONIER R~Car data together with more recent data obtained with the same instrument in 2020. Our goal is to further characterize the structures on R~Car's surface to connect them with the pulsation of the star. Sect. \ref{sec:observations} describes our observations and data reduction process. Sect. \ref{sec:analysis} presents our image analyses and results. In Sect. \ref{sec:discussion}, we discuss our results in terms of the physical parameters of the star and the angular scales obtained for the convective elements. Finally, we present our conclusions in Sect. \ref{sec:conclusions}. 

\section{Observations and data reduction \label{sec:observations}}
 
 Two datasets of observations were considered for our analysis. Both of them were taken using the 4-telescope beam combiner PIONIER. The first one consists of archival data from February of 2014; the second one corresponds to observations taken between January and March of 2020\footnote{More data were obtained for the star in 2019, however such data cover only one array configuration and are too far from the 2020 data in terms of pulsation phase to be combined together.} (see Tables \ref{tab:log_obs2014} and \ref{tab:log_obs2020}). The observations were taken at pulsation phases $\phi$ = 0.78 $\pm$ 0.02 and $\phi$ = 1.01 $\pm$ 0.1 for 2014 and 2020, respectively. To calculate the phase for the 2020 data, we followed the same methodology as for the 2014 data.
 
 All data were taken with the 4 Auxiliary Telescopes (ATs). The baselines used reached a minimum and maximum resolution of 23.5 and 1.2 mas (at $\lambda_0$=1.67 $\mu$m and in the super-resolution regime of $\lambda/2B$; being B the baseline length), respectively. The observations provided 6 spectral channels across the $H-$band between 1.516 and 1.760 $\mu$m. The scientific data of R Car were interleaved with interferometric calibrators (i.e., point-like sources with similar brightness as the science target and within a few degrees of distance) to estimate the atmospheric and instrumental response function (the calibrators are listed in Tables \ref{tab:log_obs2014} and \ref{tab:log_obs2020}). 
 
It is worth noting that for PIONIER, it is necessary to use calibration stars with brightness within $\sim$ 2 mag of the science star. This is necessary to ensure that the same instrument setup is consistently used across the entire sequence. As a consequence, the calibration stars might be significantly resolved on the longest baseline (raw visibility at about 10$\%$). However, this effect is taken into account by the pipeline, which corrects it by using the theoretical size of the calibrator spectral type. The transfer functions measured with the two different calibration stars are compatible to each other. The uncertainty on the calibration star diameters is propagated through the pipeline up to the calibrated visibilities. 
  
The calibrated interferometric observables, squared visibilities (V$^2$), and closure phases (CPs), were obtained using the \texttt{pndrs} package \citep{LeBouquin_2011}. The final calibrated observables are the average of five consecutive five-minute object observations. Considering the calibration error in the observables, the V$^2$ and CPs reached a precision of $\sigma_{\mathrm{V^2}} \sim$ 0.0025 and $\sigma_{\mathrm{CP}}\, \sim$ 1 deg, respectively. The 2014 calibrated data were obtained from the archive "Optical Interferometry Data Base" (OiDB) supported by the Jean-Marie Mariotti Center (JMMC). The interferometric observables and the uv-coverage(s) of both epochs are included in Fig\,\ref{fig:observables_2020} in the Appendix.

%--------------------------------------------------------------------
\section{Analysis and results \label{sec:analysis}}

\subsection{Diameter estimation: parametric uniform-disk model \label{sec:diameter}}

To obtain the  $H-$band angular size of R Car, we applied the geometrical model of a uniform disk (UD) to the V$^2$ data. The visibility function of this model is defined by the following equation \citep{hanbury1974angular}:

\begin{eqnarray}\label{Eq:UD}
    V_{\mathrm{UD}}(u,v) = 2F\mathrm{_r}\frac{J_1(\pi \rho \Theta_\mathrm{UD} )}{\pi  \rho \Theta_{\mathrm{UD}}}\,,
\end{eqnarray}
where $J_1$ is the first-order Bessel function; $\rho = \sqrt{u^2+ v^2}$ where $u$ and $v$ are the spatial frequencies sampled by the interferometric observations; $\Theta_\mathrm{UD}$ is the angular diameter of the uniform disk profile; and $F\mathrm{_r}$ is the scaling factor that accounts for the over-resolved flux in the observations. To fit the data, we used a Monte-Carlo Markov-Chain (MCMC) algorithm based on the \texttt{Python} package \texttt{emcee} \citep{foreman2013emcee}. We let 250 independent chains evolve for 1000 steps using the data of each one of the spectral bins independently. Finally, we averaged the best-fit UD diameters. This gave us an estimate of $\Theta_\mathrm{UD}^{2014}$ = 10.23 $\pm$ 0.05 mas (1.86 $\pm$ 0.01 au) for the 2014 epoch and $\Theta_\mathrm{UD}^{2020}$ = 13.77 $\pm$ 0.14 mas (2.51 $\pm$ 0.23 au) for the 2020 epoch.

\subsection{Asymmetries in the stellar disk: regularized image reconstruction}\label{subsec:image_reconstruction}

To characterize the asymmetric structure of R\,Car observed in the large CP variations, we reconstructed aperture-synthesis images. For the 2014 image, we adopted the methodology outlined by J. Sanchez-Bermudez and detailed in \citet{monnier20142014}. Initially, geometrical models were fitted to the interferometric observables, starting with a  single component, such as Gaussians or disks. Then, additional components were successively added to minimize the difference between the data and the models through a $\chi^2$ minimization process. Once a relatively good approximation to the data was achieved, an image was generated using the best-fit model. This best-fit model was employed as a prior image in BSMEM \citep{baron2008image}. The pixel sampling for the image reconstruction was selected to be 0.3 mas/pixel, with a pixel grid of 135 $\times$ 135 pixels. After the reconstruction process, a $\chi^2 \sim$ 6 was achieved. No further adjustments were made to this image setup, as the resulting image had been previously compared with other methods and reconstructions, confirming the validity of the observed structures.

%Since this image was already compared with other methods and reconstructions, confirming the validity of the structures observed, we did not make any further changes to the imaging setup. 

For the 2020 data, we recovered images using a combination of SQUEEZE \citep{baron2010novel} and BSMEM \citep{baron2008image}. We recovered the images using a pixel sampling of  0.3 mas/pixel, with a pixel grid of 256 $\times$ 256 pixels (which enclosures a field-of-view of $\sim$ 77 mas). First, we ran SQUEEZE using 17000 iterations over 1000 independent chains. The images were recovered independently at each of the six wavelength channels. Two regularization functions were used: Total Variation, and Compactness with hyperparameter values of one and twenty, respectively \citep[see ][ for a more complete description of each regularizer]{Sanchez-Bermudez_2018}. With this setup, we ensure the convergence of most of the chains with $\chi^2\,< $ 4. When comparing the images, we did not find significant differences across spectral channels so we averaged the chains that converged into a single image. Finally, that mean image was used as starting point for BSMEM (which uses entropy as regularizer) to smooth it without increasing the residuals in the model fitting. BSMEM converged in less than 100 iterations. Figure \ref{fig:best_mean_imgs} shows our mean best-fit images from the 2014 and 2020 data. 

\subsection{Power spectrum analysis \label{sec:psd}}

The $H-$band images of R Car's surface shown in Fig.~\ref{fig:best_mean_imgs} reveal the presence of several large bright spots. %associated with convective cells. 
The comparison between the two images reveals that those features have undergone changes over the course of seven pulsation periods from 2014 to 2020. We performed the analysis of the images' power spectral density (PSD) to estimate the characteristic size of the structures \citep{paladini2018large}. The first step is to calculate the squared modulus of the Fourier Transform of the reconstructed images. The power spectrum of an image shows the flux distribution per spatial scale or spatial frequency, the smallest frequency being the one that encloses all the normalized intensity in the image; and the highest frequency is the one that maps the normalized intensity of the most compact features. To remove the stellar disk contribution from the PSD, we set the images' pixel values below 70\% and 90\% of the intensity's peak (for the 2014 and 2020 epochs, respectively) to the median flux of the pixels above the mentioned threshold values. To characterize the weight of the power, P(k), at a given frequency, k, we computed the momentum of the power-spectrum PSD(k) = k $\times$ P(k)/P$_{\mathrm{tot}}$. 
 
\begin{figure}
    \centering
    \includegraphics[width=8.0 cm]{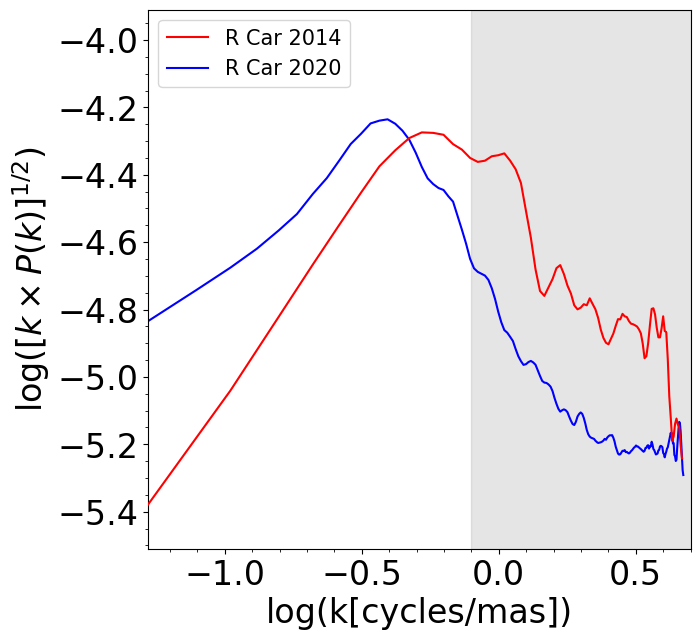}
    \caption{Radial averaged power spectra of the reconstructed images. The red and blue lines show the logarithm of the radial averaged power spectrum versus the logarithm of the spatial frequencies (in cycles/mas) of the 2014 and 2020 data, respectively. We applied a vertical shift of 0.5 to the 2020 powerspectra for better visualization.  The gray shaded area corresponds to spatial frequencies higher than the best resolution element of the interferometer.}
    \label{fig:PSDs_rcar}
\end{figure}

In order to estimate the loci of the spatial frequencies that dominate the PSDs, we averaged them radially. Fig. \ref{fig:PSDs_rcar} shows the resulting radial averaged PSDs for both epochs. We fitted a Gaussian to the PSDs' peaks to determine the dominant spatial frequencies that trace the characteristic sizes of the structures in the surface of our object. For the Gaussian fitting, we used the \texttt{Python} tool \texttt{lmfit} \citep{newville2016lmfit}. Since we only have a single image from the 2014 epoch, the reported uncertainties correspond to the standard deviation of the Gaussian fit. For the 2020 data, the model fitting was performed individually per spectral channel and the reported errors correspond to the mean of the standard deviation of the best-fit peak positions across the wavelengths. The Gaussian center was limited within the range of -0.6 < log(k) < -0.2 for the 2014 epoch and -0.7 < log(k) < -0.3 for the 2020 epoch, respectively. The measured maxima in the PSDs trace sizes of 1.82 $\pm$ 0.16 mas (4.94 $\times 10^{10}$ $\pm$ 4.89 $\times 10^{9}$ m) and 2.51 $\pm$ 0.32 mas (6.82$\times 10^{10} $ $\pm$ 9.59 $\times 10^{9} $m) for the 2014 and 2020 epochs, respectively. 

The derived sizes are bigger than the effective resolution elements of both epochs. However, since the PSD of our images can be affected by the interferometer resolution and the image reconstruction process, we quantify their impact on the determination of the measured sizes of the structures. For that purpose, we simulated two models with the following setups: (i) in the first one, we created two Uniform-Disk spots with a diameter equal to the maximum angular resolution of the interferometer (1.2 mas); (ii)  in the second one, we generated two Uniform-Disk spots with a diameter equal to the size of the surface structures derived from the 2020 image (1.8 mas). 

We extracted the interferometric observables from these two models using the interferometric u-v coverage from the 2020 epoch. We used this epoch because those data are sparser than the 2014 one, hence, we expect bigger u-v plane systematics on the reconstructed images. We performed the reconstructions following the same process as the original data. After convolving our images with the resolution of the interferometer, and extracting the characteristic (dominant) size of the structures in the images using the PSD analysis, we found that the sizes of the reconstructed spots were quite similar to the original diameters used in the models: 1.30 $\pm$ 0.01 mas and 1.81 $\pm$ 0.03 mas for the first and second simulated reconstructed images, respectively. The biggest effect is observed in the reconstructed image of the first model, with a difference $\Delta \sim$ 0.1 mas, when the simulated spot diameter is at the maximum resolution of the interferometer. However, when the simulated spot diameter equals the minimum sizes of the structures measured, the size is properly recovered with the PSD analysis. Hence, with those tests, we confirmed that the effect of the interferometer resolution and u-v coverage do not appear to have a significant effect on the measured sizes of the bright spots in the surface of R Car with the used analysis.

\section{Discussion \label{sec:discussion}}
\subsection{Sizes of the stellar disk and of the surface  structures \label{sec:sizes_granules}}

The epochs analyzed were obtained on a time baseline larger than the pulsation period of the star, hence, are part of different pulsation cycles. However, because of the stable nature of the visual light-curve (e.g.,  not sudden flux drops observed and/or secondary periods), it is reasonable to assume that  the star's properties have not changed fundamentally across the time span between our observations. Hence, throughout rest of the text, we will speak in terms of pulsation phase rather than in years of observation.

The PSD analysis shows that the diameter of the star is smaller in phase 0.78, and larger after the photometric maximum at phase 1.01. \citet{ireland2005dust} made predictions of diameter variations with pulsation for the broad $H$-band. The model predicts a shift between the size variation and the light-curve. The phase shift depends on the band of observation. In $H-$band, the object is expected to be smaller around phase 0.8, and larger before the minimum of the pulsation phase. The computed diameters agree with this trend. 

It is even more remarkable that the surface patterns are changing in size between the two pulsation phases. On average, they are smaller when the star is smaller, and larger when the star is larger. While this effect is expected from basic physics, to our knowledge, this is the first time that it is shown for the same star using images and with a quantitative analysis such as the PSD. These recovered PIONIER-VLTI images are, thus, opening a new dimension in the study of the pulsation of stars. After several years when optical interferometry measured diameter changes through the pulsation of the AGBs, the time and technology seems mature to measure the change in diameter of surface structures throughout the pulsation cycle.

\subsection{Convective pattern size(s) and their correlation with the stellar parameters}

\citet{paladini2018large} used interferometric observations to compare the structures on the surface of the S-type AGB star $\pi^1$ Gru with theoretical predictions derived from the mixing-length theory of convection \citep{ulrich1970convective}. Those authors used empirical relations of convective cell sizes obtained from 2D and 3D convective models calibrated with Solar cell sizes and applied to different stellar types  \citep{freytag1997scale, trampedach2013grid, tremblay2013granulation}. These authors found that the convective structures on the surface of $\pi^1$ Gru were consistent with the extrapolations of those convection models, suggesting that the convective process could be similar for different stellar types across the H-R diagram. 

To uncover the nature of the observed bright structures on the surface of R Car, we conducted the same experiment and compared our measurements described in Sect.\,\ref{sec:psd} with the extrapolations of the models used by \citet{paladini2018large}. First, according to \cite{freytag1997scale} (FR97 from now on), the horizontal granulation size, $x_\mathrm{g}$, is related to the characteristic pressure scale height, $H_\mathrm{p}$, by the following expression:
 
\begin{equation}
    x_\mathrm{g} = 10H_\mathrm{p}\,,
\label{eq:pressure_height}
\end{equation}

% R = 8.314 \times 10^7 erg K^-1 mol^-1, \mu = 1.3 g mol^-1 for a non ionized solar mixture with 75% Hydrogen and 25% Helium in mass, g = -0.273 for a mass = 0.74 Msun and R=195Rsun (already in table 1)
where $H_\mathrm{p} = R_{\mathrm{gas}}T_{\mathrm{eff}}/\mu\,g$, being $T_{\mathrm{eff}}$ the effective temperature of the star; $\mu$ the mean-molecular weight; $g$ is the surface gravity of the star, and $R_{\mathrm{gas}}$ is the ideal gas constant. Eq.\,(\ref{eq:pressure_height}) could be rewritten using the following logarithmic form: 

\begin{equation}
    \log(x_\mathrm{g,Freytag}) = \log(T_\mathrm{eff}) - \log(\mu) - \log(g) + 0.92\,.
    \label{eq_freytag}
\end{equation}

We calculated the predicted convective cell size of R Car using  $T_\mathrm{eff}$ = 2800 K \citep{mcdonald2012fundamental}, $R_\mathrm{gas}$ = 8.314 $\times$ 10$^7$ erg K$^{-1}$ mol$^{-1}$ and  $\mu$ = 1.3 g mol$^{-1}$ for a non-ionized solar mixture with 70\% Hydrogen, 28\% Helium and 2\% heavier elements \citep{grevesse1998standard}. To estimate $\log(g)$, we used the mass reported in \citet{takeuti2013method} of 0.87 M$_\odot$. Considering the radii  derived in Sect. \ref{sec:diameter}, we estimated $\log(g)$ values of -0.241 and -0.498 for the 2014 and 2020 epochs, respectively\footnote{ Notice that we also expect a change in temperature as a function of the radius for the two phases. This effect should not be large given that the maximum variability in the $H$ band is around 0.5 mag \citep{ireland2004pulsation}. However, with the actual data, we could not quantify this change, so we used the same effective temperature for the two data sets.}.

Second, \citet{trampedach2013grid} (TRA13 from now on) defines the following expression to calculate the cell size:

\begin{equation}
    \begin{split}
    \log(x_\mathrm{g,Trampedach}) &= (1.321 \pm 0.004)\log(T_\mathrm{eff}) \\
    &-(1.0970 \pm 0.0003)\log(g) \\
    &+(0.031 \pm 0.036).
    \end{split}
    \label{eq_trampedach}
\end{equation}
%{\Abel From this equation we have $x_\mathrm{g,Trampedach}$ = 7.06 $\times$ 10$^{10}$ m and 1.35 $\times$ 10$^{11}$ m for the 2014 and 2020 epochs, respectively.}

The same $T_\mathrm{eff}$ and $\log(g)$ values used in Eq. \ref{eq_freytag} were employed for this prescription. Third, using the CIFIST grid, \citet{ludwig2009cifist} and \citet{tremblay2013granulation} (TRE13 from now on) derived the following equation for $x_\mathrm{g}$:

\begin{equation}
    \begin{split}
        \log(x_\mathrm{g,Tremblay}) &= 1.75\log[T_\mathrm{eff}-300\log(g)]\\
        & -\log(g) + 0.05[Fe/H] - 1.87\,.
    \end{split}
    \label{eq_tremblay}
\end{equation}

\begin{figure}
    \centering
    \includegraphics[width=7.5cm]{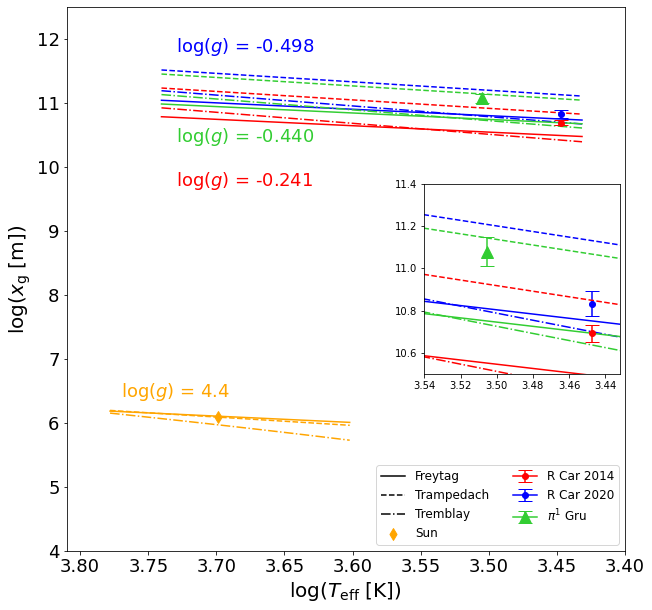}
    \caption{Logarithmic scale of the convective patterns ($x_\mathrm{g}$) versus effective temperature ($T_\mathrm{eff}$). The different lines correspond to the predictions of convective cell sizes from \citet{freytag1997scale},  \citet{trampedach2013grid} and \citet{tremblay2013granulation} (see the labels on the plot). The blue and red circles correspond to the characteristic scale of the surface structures obtained directly from the reconstructed images of R\,Car, and the green triangle shows the logarithmic value of $x_\mathrm{g}$ reported by \citet{paladini2018large} for $\pi^1$Gru. } 
    \label{fig:hgranulation_scale}
\end{figure}

In the latter formula, we assumed solar metallicity [Fe/H]=0 and the same values of $T_\mathrm{eff}$ and $\log(g)$ as in TRA13 and FR97. The logarithms of the previously derived values of $x_\mathrm{g}$ are reported in Table \ref{tab:granulation_parameters}. Figure\,\ref{fig:hgranulation_scale} also displays the logarithm of the derived $x_\mathrm{g}$ for the two reconstructed images as well as the three prescriptions. For comparison, we plot the value of $x_\mathrm{g}$ reported by \citet{paladini2018large} for $\pi^1$Gru.

In angular scales, the $x_\mathrm{g}$ from the models vary from 1 to 5 mas, being the lower and upper boundaries set by the TRE13 and TRA13 prescriptions, respectively. The $x_\mathrm{g}$ derived from our reconstructed images have sizes of 1.82 $\pm$ 0.16 mas  and 2.51 $\pm$ 0.32 mas for the 2014 and 2020 data,  respectively, situating them  in between the boundaries predicted by the models. This would imply a maximum of $\sim$ 127 cells existing on the surface of the star if we assume that all structures have a single size and can be considered circular. This is below the limit of 400 cells predicted by \citet{Schwarzschild_1975}. It is important to highlight that the derived sizes are larger than the minimum resolution of our interferometer (1.3 mas), including the effects introduced by the image reconstruction process.     
 
\begin{table}[htb]
\centering
\caption{Logarithm of the characteristic size of the patterns observed on the surface R Car. $x_\mathrm{g}$ is given in meters.}
\begin{tabular}{c@{\hskip 0.01in}@{\hskip 0.01in}c@{\hskip 0.05in}c@{\hskip 0.01in}} \hline \hline
Model      & \begin{tabular}[c]{@{}c@{}}2014\\ $\phi$ $\sim$ 0.78\end{tabular} & \begin{tabular}[c]{@{}c@{}}2020\\ $\phi$ $\sim$ 1.01\end{tabular} \\ \hline 
$\log(x_\mathrm{g,Freytag})$ & 10.49 & 10.75 \\
$\log(x_\mathrm{g,Trampedach})$ & 10.85  & 11.13   \\
$\log(x_\mathrm{g,Tremblay})$   & 10.42  & 10.70    \\
$\log(x_\mathrm{g,PSD}$)        & 10.69 $\pm$ 0.04 & 10.83 $\pm$ 0.06  \\ \hline
\end{tabular}
\label{tab:granulation_parameters}
\end{table}
%From this equation we obtain $x_g$ = 3.8 $\times 10^{10}$m for R Car by assuming a solar metallicity [Fe/H]=0. 

 Our estimations of the characteristic size of the bright structures on the surface of R Car also follow the predicted pattern from different pulsation phases, being smaller when the star is contracted and larger when the star expands. Despite this agreement with theoretical predictions, recent 3D convection models for AGB stars suggest that convection acts in a more complex way than the FR97, TRA13, and TRE13 scale relationships suggest \citep[see e.g.,][]{colom2020measuring}.  For example, lower surface gravity implies low densities both in the atmosphere and in the outer layers of the convection zone. This effect leads to large convective velocities, which imply much more violent convection when compared with, for example, the Sun. Then, large convective velocities cause large amounts of material to overshoot above the top of the convective-unstable layers. This causes the material to cool down and become optically thick due to molecular opacities. In this case, what could be observed as bright structures on the surface of the star are not the characteristic cell sizes but the material through the gaps created between the elevated optically thick material.  

This effect appears to be more important in more massive objects. For example, \citet{Chiavassa_2011a} already suggested that the radiative energy transport in RSGs is more efficient than in the Sun. RSGs are prone to higher pressure fluctuations than AGBs. As a consequence, two kinds of surface structures might appear in those kinds of stars: (i) small-scale, short-lived convective features whose sizes are consistent with the models used in \citet{paladini2018large} and in this work, but that the interferometers can not resolve due to the (on average) more distant location of RSGs \citep[as highlighted by][]{climent2020vlti} and; (ii) large-scale, long-lived features whose sizes are consistent with modified scale relationships predictions for which turbulent motions are included \citep{norris2021long, Chiavassa_2011a}.

Still, the fact that for R\,Car and $\pi^1$ Gru the measured bright spot sizes are consistent with the extrapolations from the FR97, TRA13, and TRA13 scale relationships suggests that we are observing the typical convective cell sizes, probably, because both AGB stars have similar surface gravities and progenitor masses close to solar. Thus, their convective processes are more similar to the ones of the Sun, which appears not to be the case for more massive objects. 

\section{Conclusions \label{sec:conclusions}}

We report the analysis of PIONIER-VLTI interferometric data of the M-type Mira star R\,Car. Those data consist of archival observations from 2014 and new data obtained in 2020. As a first step, we measured the stellar disk size and its change ($\Delta R \sim$ 0.3 R$_*$) at different pulsation phases using parametric model-fitting.   

The asymmetries observed on the surface of this object were characterized by using image reconstruction and power-spectrum analyses. Our findings suggest that the characteristic sizes of these structures are consistent with the extrapolations of three different models of convection. Those models establish convective cell sizes proportional to $\sim$10 times the pressure scale height in the atmosphere of the target. Our measurements of the convective structures in the surface of R Car support that these structures grow when the stellar disk expands and that they decrease when the star contracts within a pulsation cycle. This work is the first one in reporting this finding for Mira-type AGB stars. 

Those results add important evidence to the still scarce observational proofs of the role of convection in the evolution of those objects. However, state-of-the-art simulations and interferometric observations of more massive objects suggest that convection is more complex. Therefore, the FR97, TRE13, and  TRA13 scale relationships differ from predictions for more massive objects and/or when AGB stellar parameters change considerably from the ones tested with R Car. 

To constrain better the existing models, future, $H-$band aperture-synthesis (dynamic) images with a time baseline smaller than the pulsation cycle of R\,Car are, thus, necessary with the aim of describing the time evolution of the convective cell patterns and to characterize the living time of such structures. A proper comparison of those observations with hydrodynamic radiative transfer models will be mandatory. Further VLTI observations of additional AGB stars using different infrared bands (from 2.2 to 12 $\mu$m) are also part of our current and future efforts to characterize the role of convection and pulsation in shaping the evolution of O-rich Miras.

\begin{acknowledgements}
A.R.-G. acknowledges the support received through the Ph.D. scholarship (No. 760678) granted by the Mexican Council of Science CONAHCyT. A.R.-G. acknowledges the support received from the SSDF 2023-5 project "Scanning the atmosphere of R Car" from the European Southern Observatory.  J.S.-B. and A.R.-G. acknowledge the support received from the UNAM PAPIIT project IA 105023; and from the CONAHCyT “Ciencia de Frontera” project CF-2019/263975. J.S.-B. acknowledges the support received from the "Science Visitor Programme" from the European Southern Observatory and from the "Fizeau Exchange Programme" funded by WP 17 via OPTICON/RadioNET PIlot Program (grant agreement 101004719). The research leading to these results has received funding from the European Union’s Horizon 2020 research and innovation programme under Grant Agreement 101004719 (ORP). BF and SH acknowledge funding from the European Research Council (ERC) under the European Union’s Horizon 2020 research and innovation programme (Grant agreement No. 883867, project EXWINGS). MM acknowledges funding from the Programme Paris Region fellowship supported by the Région Ile-de-France. This project has received funding from the European Union’s Horizon 2020 research and innovation program under the Marie Skłodowska-Curie Grant agreement No. 945298. RS acknowledges financial support from the State Agency for Research of the Spanish MCIU through the “Center of Excellence Severo Ochoa" award for the Instituto de Astrofísica de Andalucía (SEV-2017-0709), from grant EUR2022-134031 funded by MCIN/AEI/10.13039/501100011033 and by the European Union NextGenerationEU/PRTR, and from grant  PID2022-136640NB-C21 funded by MCIN/AEI 10.13039/501100011033 and by the European Union".
This research has made use of the Jean-Marie Mariotti Center \texttt{OIFits Explorer} service \footnote{Available at \href{http://www.jmmc.fr/oifitsexplorer}{http://www.jmmc.fr/oifitsexplorer}}. We acknowledge with thanks the variable star observations from the AAVSO \footnote{Available at \href{https://www.aavso.org/}{https://www.aavso.org/}} International Database contributed by observers worldwide and used in this research.
\end{acknowledgements}

\bibliographystyle{aa}
\bibliography{bibliog.bib}

\onecolumn
\begin{appendix}
\section{PIONIER observation list}
\begin{table*}[htb!]
\centering
\caption{Log of the observations for the 2014 data. HD 80603 (Diameter = 0.94 $\pm$ 0.08 mas) and HD 81502 (Diameter = 1.23 $\pm$ 0.12 mas) were selected as calibrator stars.}
\begin{tabular}{llllllll}
\hline \hline
No. & Date        & Target   & Sta. Conf.  & $\tau_0^{min}$ - $\tau_0^{max}$ [ms]  & N. Obs. \\ \hline
    &             & HD 80603         &             &          &          \\
1   & 2014-01-22  & R Car    & D0-A1-C1-B2 & 2.18 - 8.8  & 98     \\
    &             & HD 81502         &             &            &         \\ \hline
    &             & HD 80603         &             &            &         \\
2   & 2014-01-24  & R Car    & D0-H0-G1    & 3.77 - 5.03  & 12    \\
    &             & HD 81502         &             &            &         \\ \hline
    &             & HD 80603         &             &            &         \\
3   & 2014-01-27  & R Car    & D0-G1-H0-I1 & 0.78 - 4.31  & 80    \\
    &             & HD 81502         &             &            &         \\ \hline
    &             & HD 80603         &             &            &         \\
4   & 2014-01-28  & R Car    & D0-G1-H0-I1 &1.83 - 14.92& 82       \\
    &             & HD 81502         &             &            &          \\ \hline
    &             & HD 80603         &             &            &          \\
5   & 2014-01-29  & R Car    & A1-G1-I1    & 2.58 - 10.33 & 52       \\
    &             & HD 81502         &             &            &        \\ \hline
    &             & HD 80603         &             &             &          \\
6   & 2014-01-30  & R Car    & A1-G1-J3    & 1.97 - 11.88 & 56         \\
    &             & HD 81502         &             &            &            \\ \hline
    &             & HD 80603         &             &             &          \\
7   & 2014-02-02  & R Car    & K0-A1-G1-J3 & 3.85 - 8.47 & 18         \\
    &             & HD 81502         &             &            &            \\ \hline
% 5   & 2020-02-20 & HD 71878 &             & 0.0005 & 51200         & 6.8 - 10.2 & 6        \\
%     &            & R Car    & K0-G2-D0-J3 & 0.0005 & 51200 - 25600 & 9.0 - 14.5 & 10       \\
%     &            & HD 80404 &             & 0.0005 & 51200         & 9.3 - 11.7 & 6        \\ \hline
% 6   & 2020-03-20 & HD 71878 &             & 0.0005 & 51200         & 4.4 - 5.2 & 12       \\
%     &            & R Car    & A0-B2-D0-C1 & 0.0005 & 51200 - 25600 & 5.4 - 5.8 & 10       \\
%     &            & HD 80404 &             & 0.0005 & 51200         & 5.1 - 6.4 & 6                     
\end{tabular}
\label{tab:log_obs2014}
\end{table*}

\begin{table*}[htb!]
\centering
\caption{Log of the observations for the 2020 data. HD 80404 (Diameter = 1.87 $\pm$ 0.19 mas ) and HD 71878 (Diameter = 2.95 $\pm$ 0.30 mas) were selected as calibrator stars.}
\begin{tabular}{llllll}
\hline \hline
No. & Date       & Target   & Sta. Conf.  & $\tau_0^{min}$ - $\tau_0^{max}$ [ms]  & N. Obs. \\ \hline
   &            &  HD71878 &              &  &                       \\
1   & 2020-01-20 & R Car    & A0-G1-J2-K0 & 17.0 - 21.1  & 10                     \\
    &            & HD 80404 &             &  &                       \\ \hline
   &            &  HD71878 &              &   &                       \\
2  & 2020-01-21 & R Car    & A0-G1-J2-K0  & 9.3 - 15.2 & 10                     \\
    &            & HD 80404 &             &  &                     \\ \hline
   &            & HD71878  &              &    &                      \\
3  & 2020-02-05 & R Car    & A0-B2-D0-C1  & 3.2 - 3.8 & 20                     \\
    &            &HD 80404  &             &  &                      \\ \hline
   &            & HD71878  &              &  &                      \\
4  & 2020-02-14 & R Car    & A0-B2-D0-C1  & 5.3 - 7.5 & 10                     \\
    &            &HD 80404  &             &  &                       \\ \hline
   &            & HD71878  &              &  &                       \\
5  & 2020-02-20 & R Car    & K0-G2-D0-J3  & 9.0 - 14.5 & 10                     \\
    &            &HD 80404  &             &  &                      \\ \hline
   &            & HD71878  &              &  &                      \\
6  & 2020-03-20 & R Car    & A0-B2-D0-C1  & 5.4 - 5.8 & 10                     \\
    &            & HD 80404 &             &  &     \\ \hline                
\end{tabular}
\label{tab:log_obs2020}
\end{table*}
\newpage

\section{Visual light curve and interferometric u-v coverage}
\begin{figure*}[!htb]
\centering
\includegraphics[width=14cm]{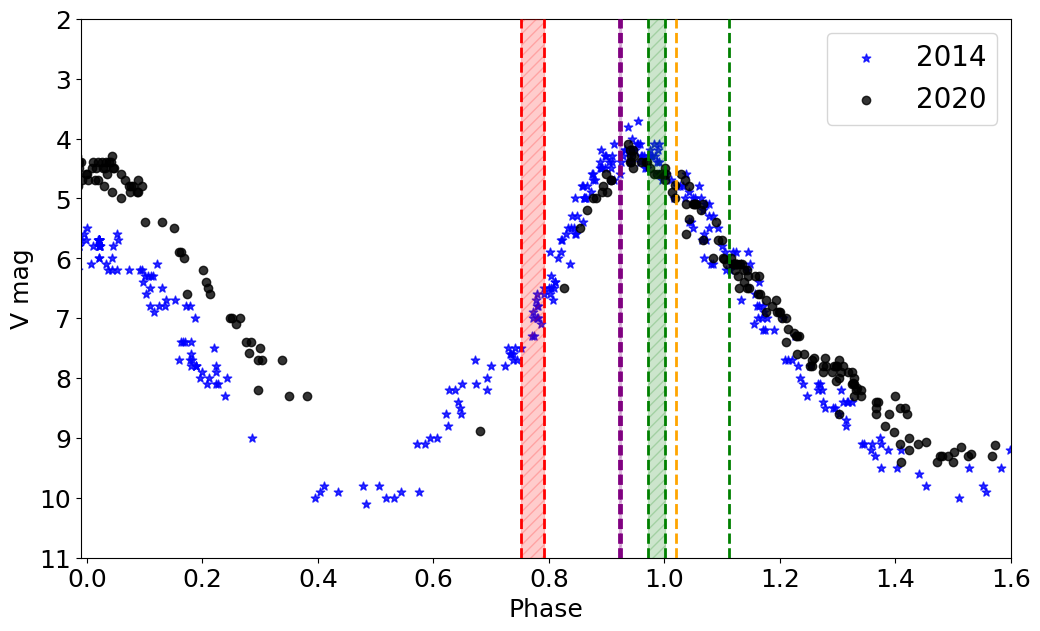}
% \subfigure[]{\includegraphics[width=9cm]{PIONIER_light_curve2014-3.png}}
% \subfigure[]{\includegraphics[width=9cm]{PIONIER_light_curve2020-3.png}}
\caption{Visual light curves of R Car obtained from the \href{https://www.aavso.org/}{AAVSO} database as a function of the pulsation phase around the dates of our interferometric observations. The blue stars and the black circles correspond to the data of the 2014 and 2020 epochs, respectively. The vertical-red shaded area indicates the phase of the 2014 interferometric observations, while the green, orange, and purple areas correspond to the 2020 ones at different station configurations: short, long, and astrometric, respectively (see Tables \ref{tab:log_obs2014} and \ref{tab:log_obs2020}).}
\label{fig:light_curves}
\end{figure*}

In this section, we present the visual light curve of R Car as a function of the phase and the interferometric u-v coverage of our observed data. 
Table \ref{tab:configs_res} displays the different observation epochs from the 2020 data set, color-coded in the same way as they appear in Fig. \ref{fig:light_curves}. We have included the maximum achievable angular resolution for each configuration, as well as the percentage of the total data covered by each epoch. Considering that the reported size of the convective structures is 1.82 $\pm$ 0.16 mas, and the highest resolution is achieved with the A0-G1-J2-K0 configuration (1.29 mas), we can conclude that configurations with the largest time interval (A0-B2-C1-D0) map larger structures, such as the stellar disk, while configurations that trace the smaller structures such as the convective structures are less temporally spaced. Therefore, by including this data set to our data, we do not expect a change in the smaller structures.

\begin{table}[htb]
\centering 
\caption{Description of our 2020 observations}
\begin{tabular}{cccccc}
\hline \hline
Color                         & Configuration & Maximum angular resolution & Dates of observation & \% of the total & Phase  \\ \hline
{\color[HTML]{6434FC} Purple} & A0-G1-J2-K0   & 1.29 mas                   & 2020-01-20           & 14.29          & 0.92 \\ 
{\color[HTML]{6434FC} Purple} & A0-G1-J2-K0   & 1.29 mas                   & 2020-01-21           & 14.29          & 0.92 \\ 
{\color[HTML]{4FD822} Green}  & A0-B2-C1-D0   & 5.2 mas                    & 2020-02-05           & 28.57           & 0.97 \\ 
{\color[HTML]{4FD822} Green}  & A0-B2-C1-D0   & 5.2 mas                    & 2020-02-14           & 14.29          & 1.00 \\ 
{\color[HTML]{FFC702} Orange} & K0-G2-D0-J3   & 1.73 mas                   & 2020-02-20           & 14.29          & 1.02 \\ 
{\color[HTML]{4FD822} Green}  & A0-B2-C1-D0   & 5.2 mas                    & 2020-03-20           & 14.29          & 1.11 \\
Total                         &               &                            &                      & 100          &     \\
\hline 
\end{tabular}
\label{tab:configs_res}
\end{table}

% {\Joel Put the dashed lines in black and, maybe, thicker. The colors used are not easily to distinguish} obtained from the \href{https://www.aavso.org/}{AAVSO} database

%\begin{figure*}
 %   \centering
  %  \subfigure[]{\includegraphics[width=8.1cm]{mcmc_fit_smallATs_feb2020_1615mic.png}}
   % \subfigure[]{\includegraphics[width=8.3cm]{mcmc_fit_smallATs_mar2020_1615mic.png}}
    %\caption{V$^2$ versus spatial frequencies of the Feb and March epochs as shown in Table \ref{tab:configs_res}. The blue lines indicate the best-fit UD models. }
    %\label{fig:v2_fit-feb_march}
%\end{figure*}

\begin{figure*}
    \centering
    \subfigure[]{\includegraphics[width=8.1cm]{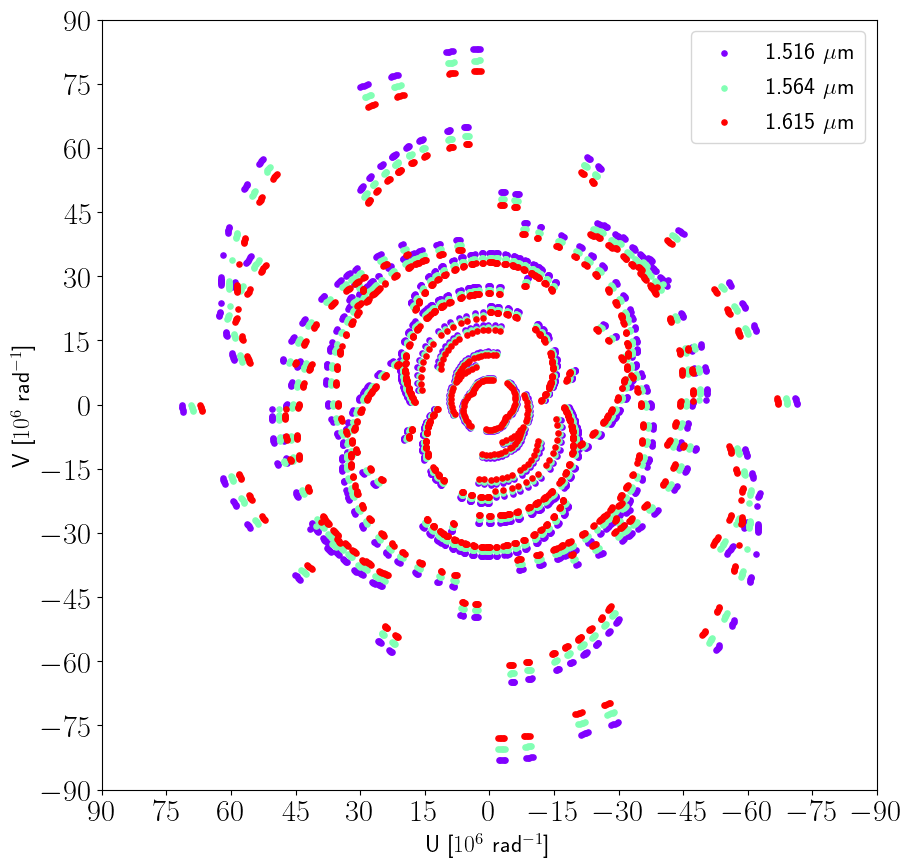}}
    \subfigure[]{\includegraphics[width=8.3cm]{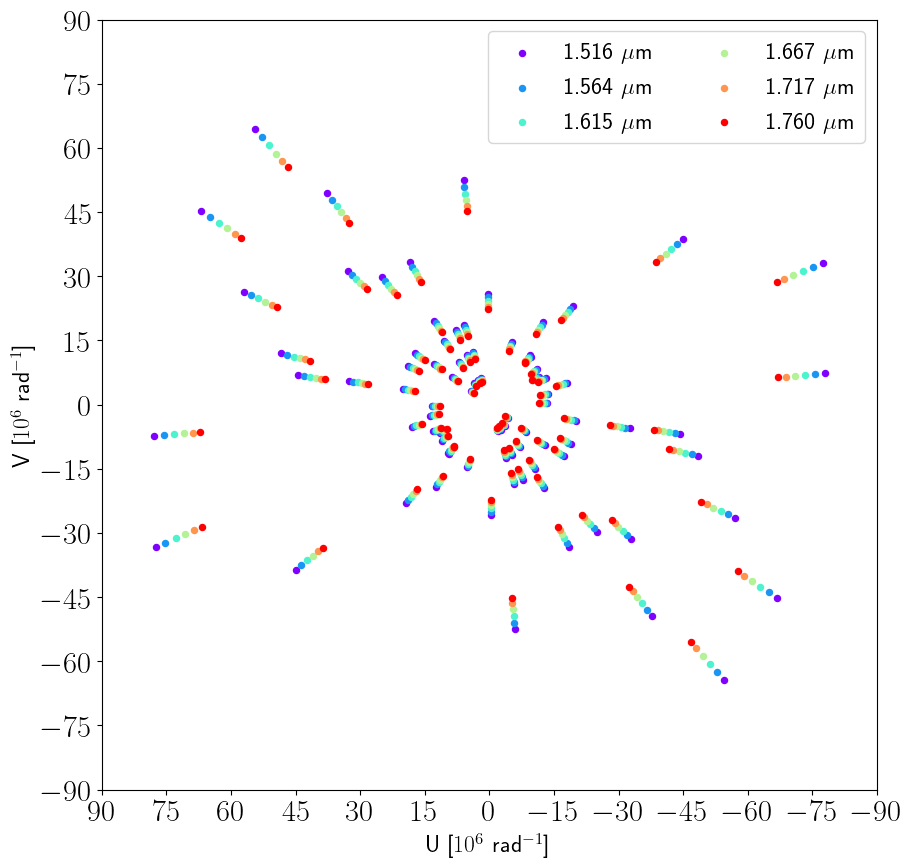}}
    \caption{PIONIER u-v coverages of the data sets used in this study. The left panel corresponds to the 2014 data set, and the right one corresponds to the 2020 one. The colors in the panels show the different wavelengths covered with the band-pass of the instrument (see the labels on the plot).}
    \label{fig:observables_2020}
\end{figure*}

\newpage
\section{Interferometric Observables}
In this section, we show the comparison between the synthetic and observed V$^2$ and CPs for our 2014 and 2020 epochs (see Sect. \ref{subsec:image_reconstruction} ). 

\begin{figure*}[!htb]
    \centering
    \includegraphics[width=15cm]{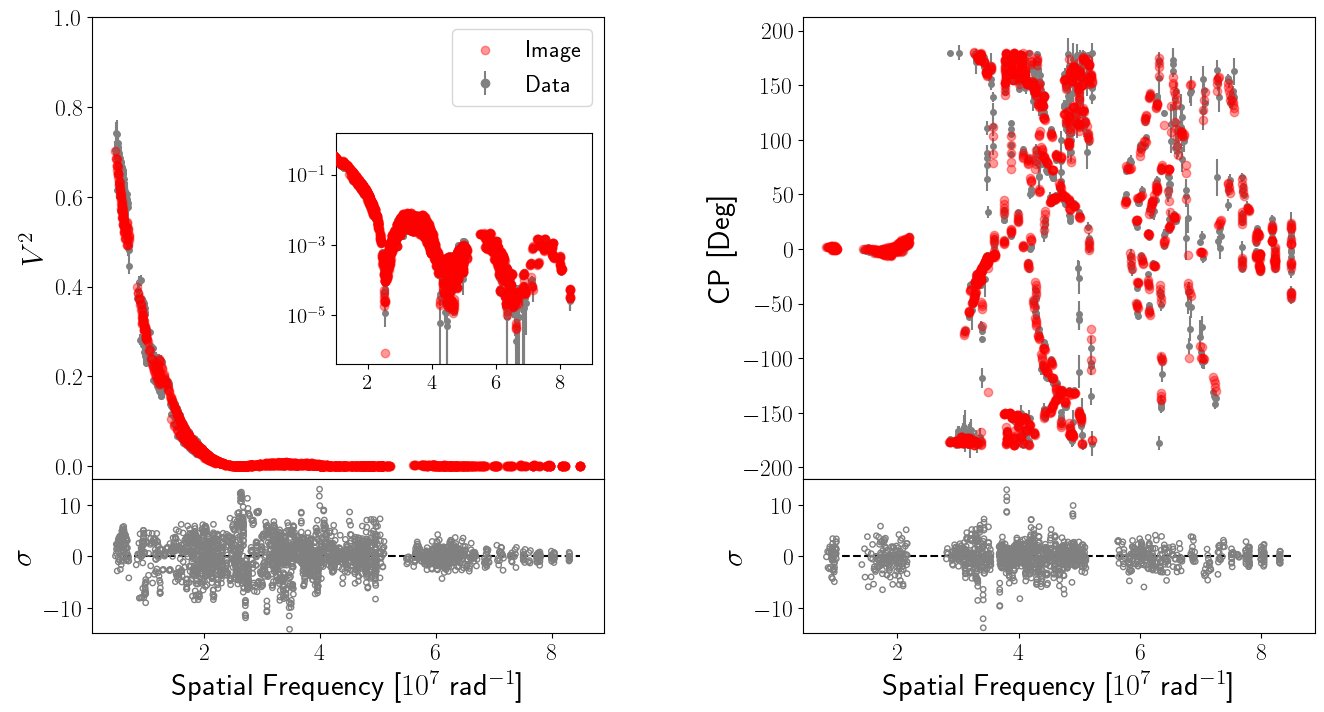}
    \caption{Observables extracted from the 2014 mean reconstructed image to the V$^2$ (left panel) and CPs (right panel) data versus spatial frequencies. The red dots correspond to the synthetic V$^2$ and CPs extracted from the raw reconstructed images, while the data are shown with gray dots.  For a better visualization, the inset shows a zoom, in logarithmic scale, of the region between 1.5 - 9 $\times$ 10$^7$ rad$^{-1}$. The lower panels show the residuals (in terms of the number of standard deviations) coming from the comparison between the data and the best-reconstructed images.}
    \label{fig:v2_t3_fit2014}
\end{figure*}

\begin{figure*}[!htb]
    \centering
    \includegraphics[width=15cm]{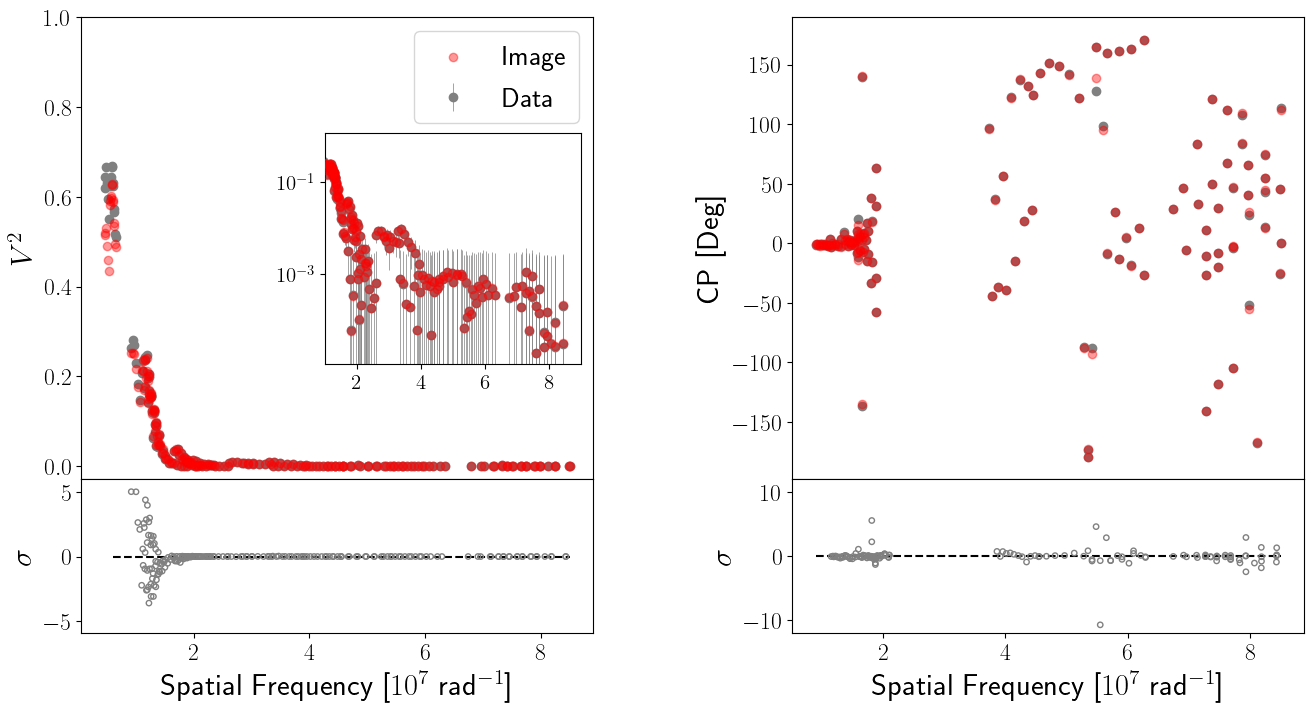}
    \caption{Best-fit observables from the 2020 reconstructed images to the V$^2$ (left panel) and CPs (right panel) data versus spatial frequencies. The red dots correspond to the synthetic V$^2$ and CPs extracted from the raw reconstructed images, while the data are shown with gray dots.  For a better visualization, the inset shows a zoom, in logarithmic scale, of the region between 1.5 - 9 $\times$ 10$^7$ rad$^{-1}$. The lower panels show the residuals (in terms of the number of standard deviations) coming from the comparison between the data and the best-reconstructed images.}
    \label{fig:v2_t3_fit2020}
\end{figure*}

\end{appendix}
 % in the main panels. The lower panels show the residuals (in terms of the number of standard deviations) coming from the comparison between the data and the best-fit reconstructed images.
\end{document}